\documentclass[]{aa}
\usepackage{graphicx}
\usepackage{rotating,subfigure,amssymb}
\usepackage{txfonts}
\usepackage{natbib}
\usepackage{subfigure}
\def \xmm {\hbox{\it XMM-Newton}}

\def \rosat {\hbox{\it ROSAT}}
\def \asca {\hbox{\it ASCA}}

\def\kT {{\rm k}T}

\def\Mv {M_{\rm 200}}

\def\rv {R_{200}}

\def\rc {r_{\rm c}}

\def \hs {\rm h_{70}}

\def \s01 {S_{0.1}}
\def \sm {$S$--$M$}
\def \st {$S$--$T$}
\def \mt {$M$--$T$}

\def \rci {r_{\rm c,in}}
\def \rcut {r_{\rm cut}}

\newcommand{\propsim}{\lower 3pt \hbox{$\, \buildrel {\textstyle
       \propto}\over {\textstyle \sim}\,$}}
\begin{document}
      \title{Structure and scaling of the entropy in nearby galaxy clusters}

      \author{G. W. Pratt$^1$, M. Arnaud$^2$ and E. Pointecouteau$^ 
{2,3}$
      }
      \offprints{G. W. Pratt, \email{gwp@mpe.mpg.de}}

      \institute{$^1$ MPE Garching, Giessenbachstra{\ss}e, 85748
        Garching, Germany  \\
                 $^2$ CEA/Saclay, Service d'Astrophysique,
                 L'Orme des Merisiers, B\^{a}t. 709,
                 91191 Gif-sur-Yvette Cedex, France \\
                 $^3$ Astrophysics, University of Oxford, Keble Road,
                 Oxford OX1 3RH, UK
                }
      \date{Received ; accepted }

\abstract{Using \xmm\ observations, we investigate the
scaling and structural properties of the ICM entropy in a sample of 10
nearby ($z < 0.2$) relaxed galaxy clusters in the temperature range 2-9
keV. We derive the local entropy-temperature (\st) relation at $R =
0.1, 0.2, 0.3$ and $0.5\rv$. The logarithmic slope of the relation is
the same within the $1\sigma$ error at all scaled radii. However, the
intrinsic dispersion about the best fitting relation is significantly
higher at $0.1\rv$. The slope is $0.64\pm0.11$ at $0.3\,\rv$, in
excellent agreement with previous work.  We also investigate the
entropy-mass relation at density contrasts $\delta=5000, 2500$ and
1000. We find a shallower slope than that expected in simple
self-similar models, which is in agreement with the observed
empirically-determined entropy-temperature and mass-temperature
scaling. The dispersion is smaller than for the $S$--$T$ relation.
Once scaled appropriately, the entropy profiles appear 
similar beyond $\sim 0.1\rv$, with an intrinsic dispersion of $\sim
15$ per cent and a shape consistent with 
gravitational heating ($S(r) \propsim r^{1.1}$).  However, the scatter
in scaled entropy profiles increases with smaller scaled radius, to
more than $60$ per cent at $R \lesssim 0.05 \rv$. Our results are in
qualitative agreement with models which boost entropy production at
the accretion shock. However, localised entropy modification may be needed
to explain the dispersion in the inner regions.
 }

      \authorrunning{G.W. Pratt et al.}
      \titlerunning{An update on galaxy cluster entropy scaling} \maketitle
%
%

\section{Introduction}

X-ray observations of the hot, gaseous intracluster medium (ICM) have
been telling us for well over a decade that physical processes other  
than
gravity are acting to modify the properties of the cluster population.
An understanding of the source(s) of this modification is of great
importance for our understanding of cluster formation and evolution,
and is essential for the use of clusters as precision cosmological
probes.

Entropy is important because, together with the shape of the potential
well, it is the quantity which dictates the observed X-ray properties
of the ICM in galaxy clusters \citep{voit05}. The intracluster entropy
is generated in shocks as gas is drawn into the gravitational potential
of the cluster halo \citep{tn01, borg01,
voit02, voit03, voit05, borg05}, thus it reflects the accretion
history of the ICM. However, the entropy distribution also preserves
key information regarding the influence of non-gravitational
processes.

X-ray measurements of the entropy\footnote{Throughout this paper the
entropy is defined as $S = \kT n_e^{-2/3}$. This quantity is related
to the true thermodynamic entropy by a logarithm and an additive
constant.} at $0.1 \rv$ (hereafter $S_{0.1}$) showed that the entropy
of the coolest systems is considerably higher than that available from
gravitational collapse \citep{pcn99,lpc00}, and that the
entropy-temperature (\st) relation is shallower than expected
\citep{psf03}. More recent spatially resolved entropy profiles
indicate that the entropy is higher {\it throughout} the ICM, and
that, outside the core regions, entropy profiles are structurally
similar \citep{psf03,vp03,pa03,pa05,piff}. At the same time the
scatter in $ \s01 $ at a given temperature can be up to a factor of
three \citep{psf03}.

Possible entropy modification mechanisms have historically included
preheating, where the gas has been heated before being accreted into
the potential well, by early supernovae and/or AGN activity
\citep[e.g.,][]{kaiser91, evrard91,valageas}, internal heating after
accretion \citep[e.g.,][]{me94}, and cooling
\citep[e.g.,][]{pea00}. The lack of isentropic core entropy profiles
in groups and poor clusters has shown that simple preheating is
unlikely to be the sole explanation of the observations
\citep{psf03,pa03,pa05}. Since cooling-only models generally predict a
higher stellar mass fraction than observed \citep[e.g.,][]{muan}, attention
is now focussing on the interplay between cooling and
feedback. Further high quality observations are needed in order to
distinguish between these different entropy modification mechanisms.

\begin{figure*}[ht]
\begin{centering}
\includegraphics[scale=1.,angle=0,keepaspectratio,width=0.97 
\columnwidth]{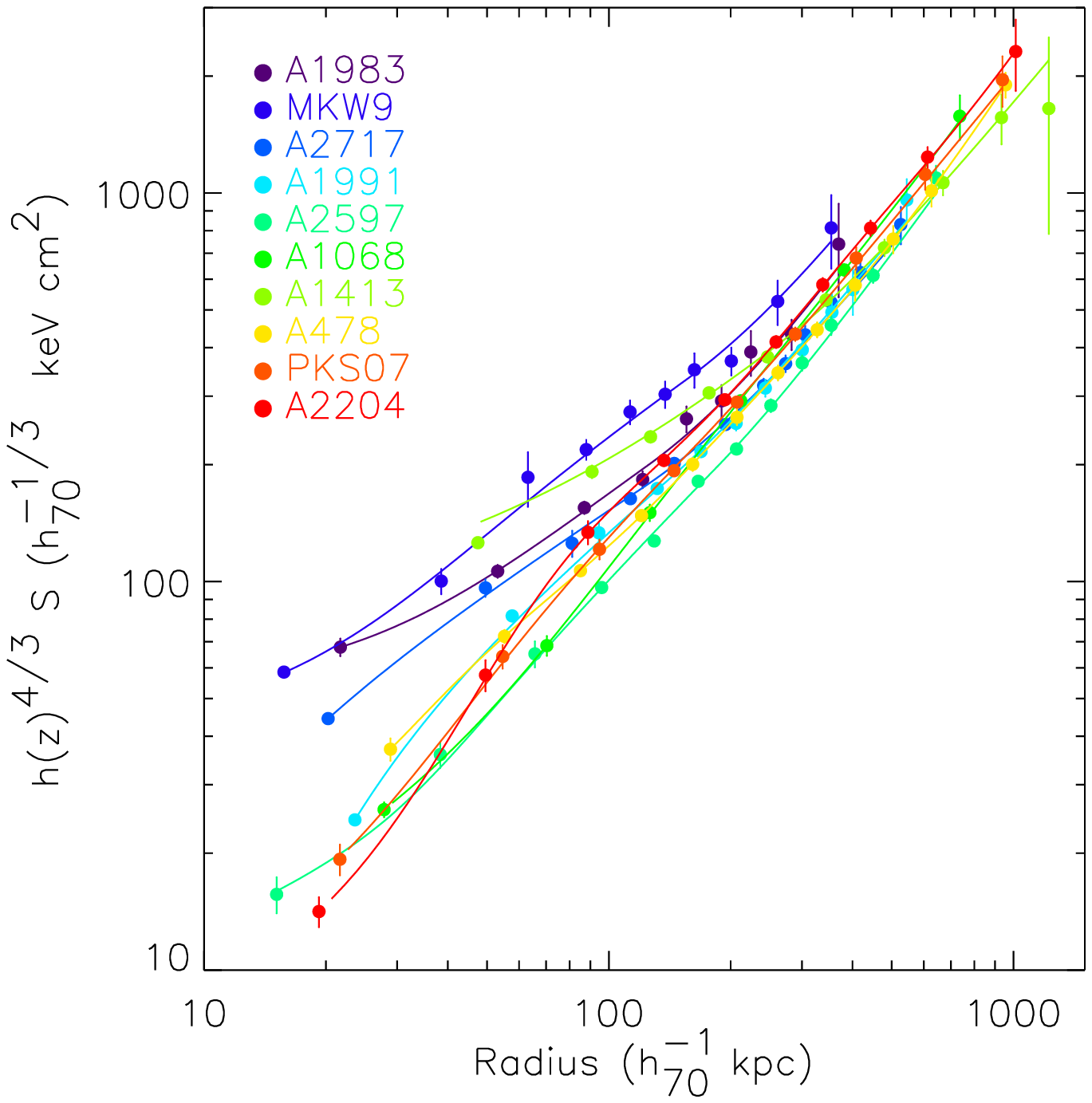}
\hfill
\includegraphics[scale=1.,angle=0,keepaspectratio,width=
\columnwidth]{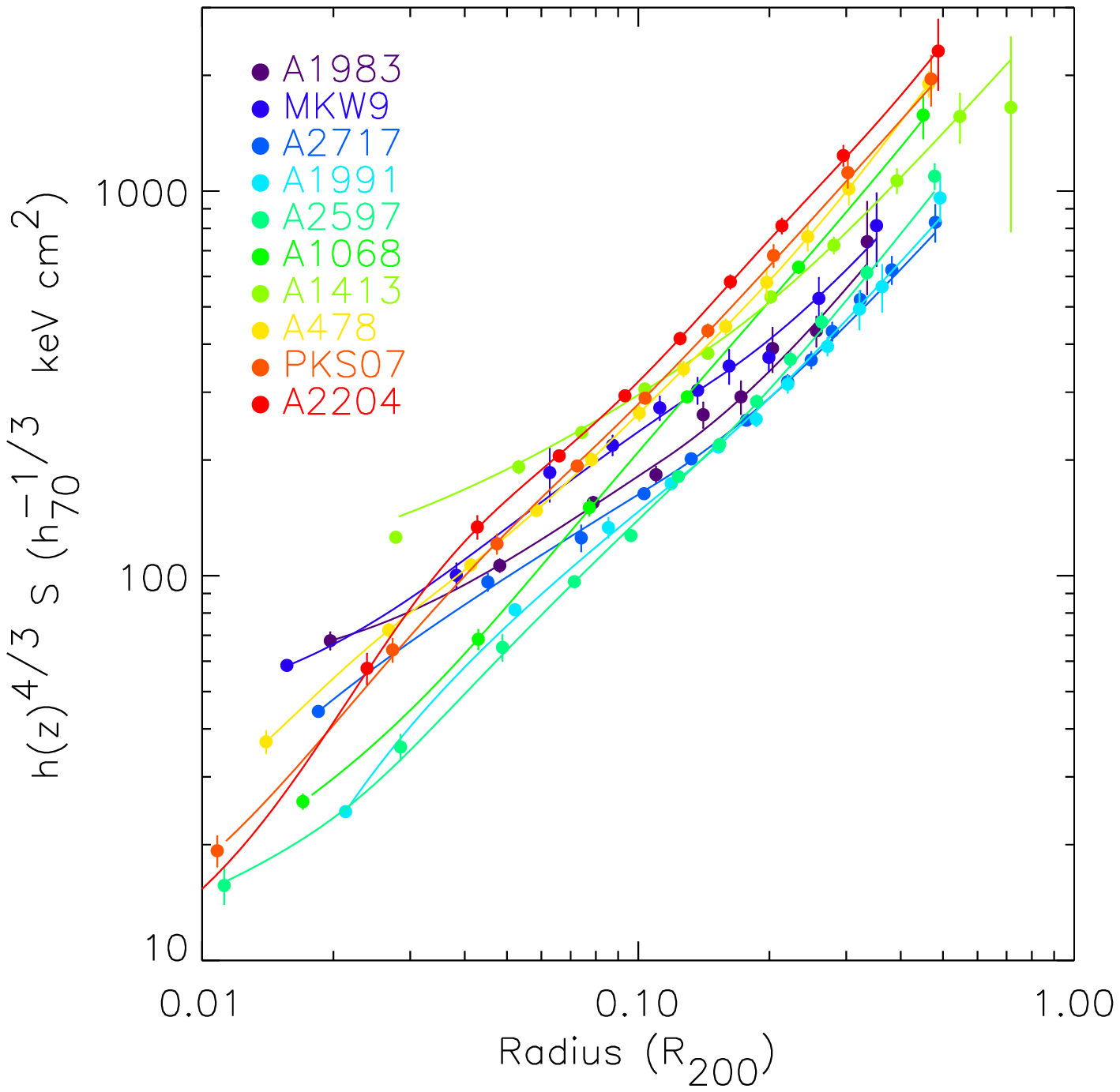}
\caption{{\footnotesize Cluster entropy profiles obtained from the
deprojected, PSF corrected temperature profiles and the best fitting
analytical model for the gas density. Solid lines, included to
improve visibility, are entropy profiles obtained from analytic model
fits to the temperature and density information. On the left, the
profiles are plotted in physical units ($h_{70}^{-1}$ kpc); on the
right, they are plotted in units of $\rv$.}}\label{fig:rawprof}
\end{centering}
\end{figure*}

The present paper is the third in a series based on \xmm\
observations of ten clusters in the temperature range from 2 keV to 9
keV. In \citet*[][hereafter Paper~I]{pap}, we investigated the shape
and scaling properties of the dark matter distribution, and in
\citet*[][hereafter Paper~II]{app} we examined the relation between
the total mass and the X-ray temperature. In the present paper we
extend the work of \citet{pa05} by exploring the
entropy scaling properties over a larger temperature/mass
range.

All results given below were calculated assuming a $\Lambda$CDM
cosmology with $\Omega_m=0.3$ and $\Omega_\Lambda=0.7$ and $H_0 =
70$~km~s$^{-1}$~Mpc$^{-1}$. Unless otherwise stated, errors are
given at the $68$ per cent confidence level.

\section{Sample and analysis}\label{sec:sample}

\begin{table}
\begin{minipage}{\columnwidth}
\caption{{\footnotesize Basic cluster data.}}\label{tab:obs}
\centering
\begin{tabular}{l l l}
\hline
\hline
\multicolumn{1}{l}{Cluster} & \multicolumn{1}{l}{z} &
\multicolumn{1}{l}{T} \\
\multicolumn{1}{l}{ } & \multicolumn{1}{l}{} &
\multicolumn{1}{l}{(keV)} \\
\hline
A1983 & 0.0442 & $2.18\pm0.09$ \\
A2717 & 0.0498 & $2.56\pm0.06$ \\
MKW9 &  0.0382 & $2.43\pm0.24$ \\
A1991 & 0.0586 & $2.71\pm0.07$ \\
A2597 & 0.0852 & $3.67\pm0.09$ \\
A1068 & 0.1375 & $4.67\pm0.11$ \\
A1413 & 0.1427 & $6.62\pm0.14$ \\
A478 &  0.0881 & $7.05\pm0.12$ \\
PKS0745 &       0.1028 & $7.97\pm0.28$ \\
A2204 & 0.1523 & $8.26\pm0.22$ \\

\hline
\end{tabular}
\end{minipage}
\end{table}

The sample (Table~\ref{tab:obs}) comprises 10 systems ranging in
temperature from 2 keV to $\sim8.5$ keV.  In Paper I, spatially
resolved temperature and density data were used to calculate
gravitating mass profiles. These were which then fitted with an NFW
model, yielding values for $\Mv$ and $ \rv$ ('virial' mass and radius
corresponding to a density contrast\footnote{The density contrast,
$\delta = 3 M(<r)/4\pi r^{3} \rho_{\rm c}(z)$, where $\rho_{\rm c}(z)
= h^{2}(z) 3 H_0^2 / 8 \pi G$, and $h^{2}(z) = \Omega_{\rm M}(1 +
z)^3 + \Omega_\Lambda$.}  of 200).  The mean temperature of
each system, $T_{\rm spec}$, needed to investigate the scaling
properties of the profiles, was estimated from fits to a spectrum
built from all events in the region $0.1\rv\,\leq r \leq\,0.5\rv$ (see
the discussion in Paper~II). The data reduction steps are described in
detail in Paper~I.

We calculated the entropy profile of each system using the
deprojected, PSF corrected temperature profile and an analytical model
for the gas density derived in Paper I. Analytic gas density model
parameters are listed in Appendix~A. In Fig~\ref{fig:rawprof}, the raw
entropy profiles are shown in physical units ($\hs^{-1}$ kpc), and in
terms of the measured virial radius, $\rv$.  All profiles increase
monotonically with radius and, while the slope of the profile becomes
shallower towards the centre in some of the clusters, none shows an
isentropic core.


\section{Entropy scaling properties}

\subsection{Introduction}

In a self-similar cluster population expected from simple
gravitational collapse, the scaled profiles of any
physical quantity coincide, and thus measures of these quantities at
any {\it scaled\/} radius should correlate with global quantities
such as the mean temperature, or the virial mass.  In this standard
model, the entropy is expected to scale as $S  
\propto h(z)^{-4/3}T$ or $S\propto h(z)^{-2/3}M^{2/3}$.

In purely theoretical terms, the most fundamental characteristic of a
cluster is its mass.  Previous investigations of the entropy scaling
relations were hampered in this respect because they did not have
accurate mass information. The present observations, which have
excellent mass data (Papers~I and~II), thus represent an ideal
opportunity to examine the entropy-mass relation.  In addition, the
spatial resolution and radial coverage in the entropy profiles are
significantly improved, as compared with previous \rosat/\asca\
studies \citep{psf03}. In this Section, we investigate both the
entropy-temperature ($h(z)^{4/3} S$--$T_{\rm spec}$) and the
entropy-mass ($h(z)^{2/3}S$--$\Mv$) relations.  We derive the slope
and normalisation of each relation, at various scaled radius/density
contrasts. We also estimate the raw, statistical and intrinsic scatter
about the relations.

\begin{table}
\begin{minipage}{\columnwidth}
\caption{{\footnotesize The $S$--$T$ relation. Data were fitted with a
     power-law of the form $h(z)^{4/3}~S = A \times (\kT/{\rm 5
       keV})^{\alpha}$, where $\kT$ is the overall spectroscopic
     temperature in the $0.1-0.5~\rv$ region. Errors
     in entropy and temperature are taken into account. Results are  
given for the WLS, BCES, and WLSS regression methods  (see text).  
Raw, statistical and intrinsic scatter about the best fitting  
relation in the log-log plane  are given in the last columns; the
numbers in parentheses indicate the percentage intrinsic scatter.}}\label 
{tab:strel}
\centering
\begin{tabular}{l r l l l l}
\hline
\hline

\multicolumn{1}{c}{Radius} &
\multicolumn{1}{c}{$A$} &
\multicolumn{1}{c}{$\alpha$} & \multicolumn{3}{c}{$\sigma_{log}$} \\
\cline{4-6}
\multicolumn{1}{c}{$\rv$} &
\multicolumn{1}{c}{keV cm$^{-2}$} &
\multicolumn{1}{l}{ } &  \multicolumn{1}{l}{raw} &
\multicolumn{1}{l}{stat} & \multicolumn{1}{l}{int} \\

\hline

WLS\\
\cline{1-1}
0.1   & $227 \pm 5$ & $0.58\pm0.05$ & 0.079 & 0.030 & 0.073\\
0.2   & $478 \pm 12$ & $0.73\pm0.06$ & 0.058 & 0.035 & 0.047  \\
0.3   & $770 \pm 24$ & $0.71\pm0.07$ & 0.074 & 0.043 & 0.060 \\
0.5   & $1510 \pm 90$ & $0.68\pm0.12$ & 0.070 & 0.078 & - \\

BCES\\
\cline{1-1}

0.1   & $230 \pm 17$ & $0.49\pm0.15$ & 0.082 & 0.030 & 0.076 $ 
\scriptstyle (19\%)$  \\
0.2   & $485 \pm 22$ & $0.62\pm0.11$ & 0.063 & 0.034 & 0.052 $ 
\scriptstyle(13\%)$\\
0.3   & $798 \pm 44$ & $0.64\pm0.11$ & 0.078 & 0.043 & 0.065 $ 
\scriptstyle(16\%)$ \\
0.5   & $1560 \pm 83$ & $0.62\pm0.08$ & 0.074 & 0.078 & - \\

WLSS\\
\cline{1-1}
0.1   & $229 \pm 16$ & $0.47\pm0.14$ & 0.083 & 0.030 & 0.077 \\
0.2   & $480 \pm 23$ & $0.67\pm0.10$ & 0.059 & 0.035 & 0.048 \\
0.3   & $786 \pm 43$ & $0.69\pm0.12$ & 0.075 & 0.043 &0.061  \\
0.5   & $1510 \pm 90$ & $0.68\pm0.12$ & 0.070 & 0.078 & -  \\

\hline
\end{tabular}
\end{minipage}
\end{table}

\begin{figure}[t]
\begin{centering}
\includegraphics[scale=1.,angle=0,keepaspectratio,width=\columnwidth] 
{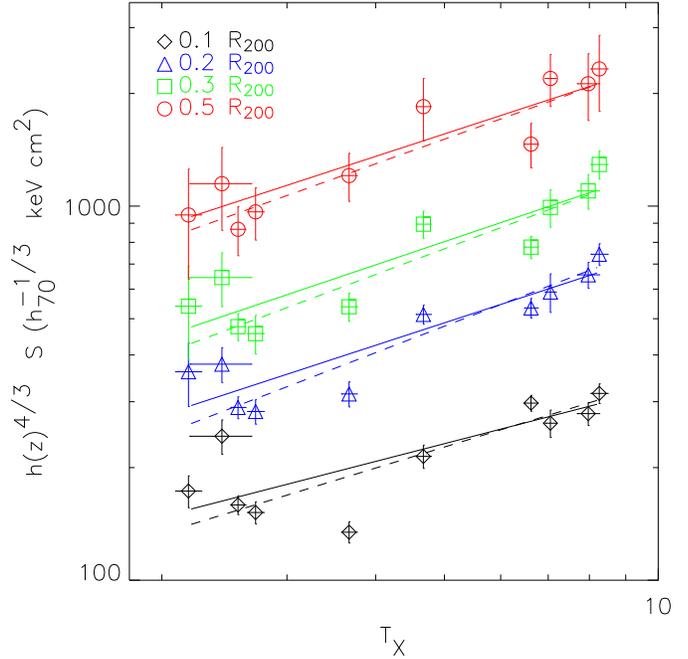}
\caption{{\footnotesize The $S$-$T$ relation measured from a sample of
     10 clusters covering a temperature range from 2 to 9 keV. The
     $S$-$T$ relation is shown for different fractions of
     $\rv$. Measurements are plotted with error bars. At each radius,
     the best-fitting power-law relation,
     derived taking account the errors in entropy and temperature,
     is overplotted. Solid line: BCES regression method; dashed line:  
WLS
     method; slopes and intercepts are given in
     Table~\ref{tab:strel}.}}\label{fig:strel}
\end{centering}
\end{figure}

\subsection{Method}

The entropy estimates at a given scaled radius ($x=r/\rv$) or density  
contrast ($\delta$) have to be calculated taking into account the
uncertainties on the temperature, the analytical gas density model,
and the measurement of the scaled radius/density contrast itself.
We thus
estimated the entropy at a given $x$ or $\delta$, and the associated
errors, using a Monte Carlo method. The scaled radius $r(x)$/$R_ 
\delta$ values and
corresponding uncertainties were estimated from NFW-type modelling of
the mass profile (see Papers~I and~II for details), and subsequently
randomised assuming a Gaussian distribution with sigma equal to the $1
\sigma$ error. The temperature profile was then similarly randomised
taking into 
account the observed uncertainties, and the temperature estimated at
the relevant (randomised) $r(x)$/$R_\delta$ using spline 
interpolation. The entropy was then calculated in the usual manner, $S
= \kT / n_e^{2/3}$, including an additional $5$ per cent systematic
error on $n_e$ (corresponding to the uncertainties in the surface
brightness profile modelling; see Paper~I). This procedure was
undertaken 1000 times for each cluster; the final entropy value at
each $x$/ $\delta$ was calculated from the mean and standard deviation
of the randomised values at that radius.

The temperature profiles of all systems except A1983 and MKW9 are
detected up to $\sim 0.5\,\rv$ (the scaled radius/density contrast of
maximum detection of A1983 and MKW9 are 
$0.38\rv$/1455$\delta$ and $0.41\rv/1401\delta$, respectively).
The temperature profiles of these clusters had thus to be  
extrapolated to
$0.5\rv$/$1000\delta$. To avoid spuriously high/low temperatures, we
only allowed extrapolations within $\pm 1.5$ times the mean temperature
of the cluster (well within the range of temperature gradients
observed to date, e.g., \citealt{app,vikh}).

\subsection{The $S$--$T$ relation}
\label{sec:strel}

Our data have the required temperature coverage to investigate the
entropy-temperature relation. For easier comparison with previous
work, we first investigated the $S$--$T$
relation at different fractions of the virial radius, viz., 0.1, 0.2,
0.3, and $0.5~\rv$.  We fitted a power law model of the form

\begin{equation}
h(z)^{4/3}~S_{\rm x} = A [T_{\rm spec}/{\rm 5~keV}]^\alpha
\end{equation}

\noindent to the data, $T_{\rm spec}$ being defined as in
Sect.~\ref{sec:sample}.  We chose a pivot point of 5 keV for
consistency with the $M$--$T$ relation of \citet{app}.  The fit was
performed using linear regression in the log-log plane, taking into
account the errors on both variables. We have also computed the raw
and intrinsic scatter about the best fitting relations in the log-log
plane.  To estimate the raw scatter, we used the vertical distances to
the regression line, weighted by the error. The intrinsic scatter was
computed from the quadratic difference between the raw scatter and the
scatter expected from the statistical errors. The resulting values are
given in Table~\ref{tab:strel}.

In undertaking the fits, we first used the classical weighted least
square method, WLS ($\chi^ {2}$ estimator) implemented in the the
routine FITEXY from numerical recipes \citep{numrec}. This regression
method is strictly valid only if the intrinsic scatter is negligible
as compared to the statistical scatter. This is generally not the case
(see Table~\ref{tab:strel}). We thus also considered the orthogonal
BCES method \citep{akritas96}, with bootstrap resampling. This is the
least-biased regression method when both measurement errors and
intrinsic scatter are present. However, no error weighting is
performed on individual data points.  We thus also considered a
variation of the WLS method, where a constant term, corresponding to
the intrinsic scatter, is added quadratically to the $
\log(S)$ error (WLSS  
method). The best fit is determined such that the reduced $\chi^{2}$,
after minimization over $A$ and $\alpha$, is $\chi^{2}_{\rm red}=1$.

The best fitting slopes and intercepts for the different methods are
listed in Table~\ref{tab:strel}. Figure~\ref{fig:strel} shows the data
and the best-fitting power law relation for each radius under
consideration, obtained using the standard WLS and BCES methods.  The
three methods give best fitting parameters consistent within the
$1\sigma$ errors.  However, the errors are underestimated with the WLS
method, especially at low radii. This is a consequence of the high
$\chi^{2}$ value obtained, reflecting the high intrinsic scatter at
small scaled radius (e.g. $\chi^{2} = 54$ for 8 d.o.f at
$0.1~\rv$). In this case the standard $\Delta\chi^{2}=1$ criteria to
derive parameter errors is not valid.  The BCES and WLSS give results
in excellent agreement with each other, both in terms of best-fitting
values and associated uncertainties. In the following, for easier
comparison with previous work, we will refer to results obtained using
the commonly used BCES method.

The slope of the entropy-temperature relation is incompatible with the
standard self-similar prediction at all radii at which we have measured it,
confirming the results of \citet{psf03}.  Our best-fitting
$S_{0.1}$--$T$ slope ($\alpha=0.49\pm0.15$) is slightly shallower than
that found by
\citeauthor{psf03} from unweighted orthogonal  
regression on individual data points ($\alpha=0.65\pm0.05$). A better
agreement is observed with the slope derived by these authors from
data grouped in temperature bins ($\alpha=0.57\pm0.04$).  However, the
slopes are consistent once uncertainties are taken into account.

Beyond $r \geq 0.2\,\rv$, the slope remains remarkably stable. At $0.3
\rv$, our gas density and temperature measurements are well
constrained, and do not suffer potential systematic problems connected
to the correction for PSF and projection effects. At this radius, we
are well outside the strong cooling cores found in the majority of the
objects in our sample. In addition, we note that there are no
significant temperature gradients at this radius in the high
resolution temperature profiles produced by \citet{vikh} from {\it
Chandra} observations. For this reason we consider the measurements at
$0.3\rv$ to be the most reliable. The slope of the $S_{0.3}$-$T$
relation, obtained using the BCES method, is $S_{0.3} \propto
T^{0.64\pm0.11}$, in excellent agreement with that found by
\citet[][]{psf03}.

Figure~\ref{fig:strel} shows that there is noticeable scatter about
the $S_{0.1}$--$T$ relation. Table~\ref{tab:strel} makes clear that
the scatter is reduced at larger scaled radius.  The intrinsic scatter  
remains the dominant contributor to the
dispersion in all relations, except at $0.5~\rv$.


\subsection{The $S$--$M$ relation}

\begin{table}[t]
\begin{minipage}{\columnwidth}
\caption{{\footnotesize The $S$--$M$ relation. Data were fitted with a
     power-law model of the form $h(z)^{2/3}~S_{\delta} = B_{\delta}  
\times
     (M_{200}/5.3 \times 10^{14} M_{\odot})^{\beta}$, where $M_{200}$
     is the total mass obtained from NFW model fits to the measured
     mass profiles. Best-fitting relations are given for the BCES
     estimator, taking into account errors in mass and entropy. Results
     are given for the full sample, and for the sample excluding MKW9,
     which  is a significant outlier. The numbers in parenthesis
     indicate the percentage intrinsic scatter.}}\label{tab:smrel}
\centering
\begin{tabular}{l r l l l l}
\hline
\hline

\multicolumn{1}{c}{$\delta$} &
\multicolumn{1}{c}{$B$} &
\multicolumn{1}{c}{$\beta$} & \multicolumn{3}{c}{$\sigma_{log}$} \\
\cline{4-6}
\multicolumn{1}{c}{} &
\multicolumn{1}{c}{keV cm$^{-2}$} &
\multicolumn{1}{l}{ } &  \multicolumn{1}{l}{raw} &
\multicolumn{1}{l}{stat} & \multicolumn{1}{l}{int} \\

\hline
Full \\
sample  \\
\cline{1-1}
5000    & $471\pm18$ & $0.36\pm0.10$ & 0.058 & 0.034 & 0.046 $ 
\scriptstyle (11\%)$ \\
2500    & $765\pm30$ & $0.37\pm0.10$ & 0.059 & 0.041 & 0.043 $ 
\scriptstyle (10\%)$ \\
1000    & $1460\pm47$ & $0.36\pm0.06$ & 0.059 & 0.065 & - \\

Excl. \\
MKW9 \\
\cline{1-1}

5000    & $459\pm37$ & $0.43\pm0.10$ & 0.039 & 0.035 & 0.017 $ 
\scriptstyle (4\%)$\\
2500    & $741\pm43$ & $0.44\pm0.08$ & 0.035 & 0.041 & - \\
1000    & $1430\pm46$ & $0.41\pm0.04$ & 0.042 & 0.063 & - \\

\hline
\end{tabular}
\end{minipage}
\end{table}

\begin{figure}[t]
\begin{centering}
\includegraphics[scale=1.,angle=0,keepaspectratio,width=\columnwidth] 
{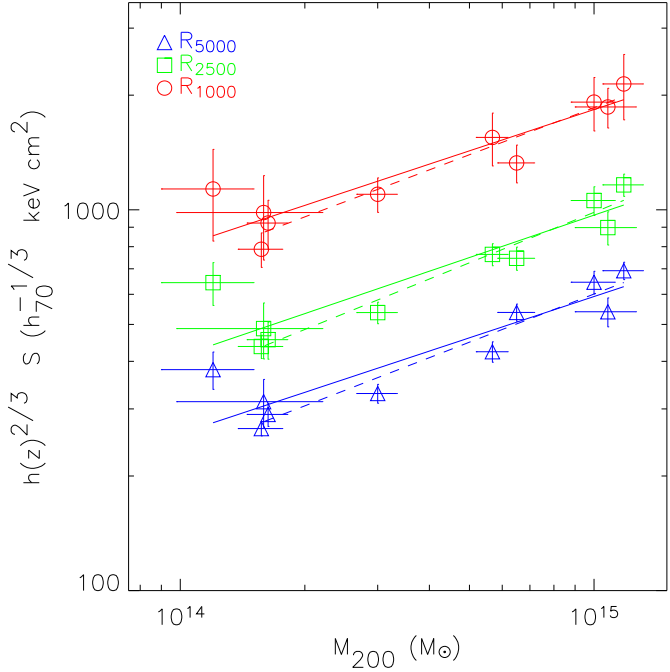}
\caption{{\footnotesize The $S$--$M$ relation measured at various
     density contrasts. $M_{200}$ values were measured from NFW  
        model 
     fits to the measured mass profiles (Papers~I and II). Symbols with
     error bars represent the measurements. Lines show the
     best-fitting power-law $S$--$M$ relation: solid: full sample;
     dashed: excluding MKW9.  Slopes and
     intercepts are given in Table~\ref{tab:smrel}.}}\label{fig:smrel}
\end{centering}
\end{figure}

In investigating the $S$--$M$ relation, it is more logical to work in
terms of density contrast $\delta$.  We used $\delta =
5000$, 2500, and 1000, which correspond to average
fractions of $0.20\pm0.01~\rv$, $0.29\pm0.02~\rv$ and $0.47\pm
0.02~\rv$ for the present sample, and fitted a power law of the form

\begin{equation}
h(z)^{2/3} S_\delta = B_\delta [M_{200}/5.3 \times 10^{14}
M_\odot]^\beta
\end{equation}

\noindent to the data. The pivot point of $5.3 \times 10^{14}
M_\odot$ corresponds to a temperature of 5 keV using the best-fitting
$M$--$T$ relation of \citet{app}. In Fig~\ref{fig:smrel} we show the
data and the 
best-fitting power law relation for each density contrast under
consideration. Since MKW9 appears to be a clear outlier, we also
fitted the relation excluding this cluster. The best fitting slopes
and intercepts for the \sm\ relation, with and without MKW9, are
listed in Table~\ref{tab:smrel}.

The best-fitting $S$--$M$ relation is shallower
than the prediction from standard self-similar models. The results for the
full sample are entirely consistent with the observed \st\ and \mt\
relations (Sect.~\ref{sec:strel} and Paper~II). Excluding MKW9, the
slope is slightly, although not significantly, steeper.

As can clearly be seen in Fig.~\ref{fig:smrel}, and as quantified in
Table~\ref{tab:smrel}, the intrinsic dispersion about the
\sm\ relations is less than that for the \st\
relations. This is particularly true for the sample excluding
MKW9. This is not an artifact of the larger statistical mass errors
compared to those on the temperature: the dispersion is similarly
smaller for the raw scatter.  Table~\ref{tab:smrel} also indicates
that, excluding MKW9, the dispersion is now dominated by the
statistical scatter at all density contrasts.

\section{Scaled entropy profiles}

We have confirmed that the $S$--$T$ relation is shallower than
expected in the standard self-similar model, and we have shown that
this is also true for the $S$--$M$ relation.  In this Section, we
examine the structural properties of the entropy profiles. We do this
by looking at the entropy profiles once scaled by the best fitting
relations.

\subsection{Scaled entropy versus $r/R_{200}$}

\begin{figure}[t]
\begin{centering}
\includegraphics[scale=1.,angle=0,keepaspectratio,width=\columnwidth] 
{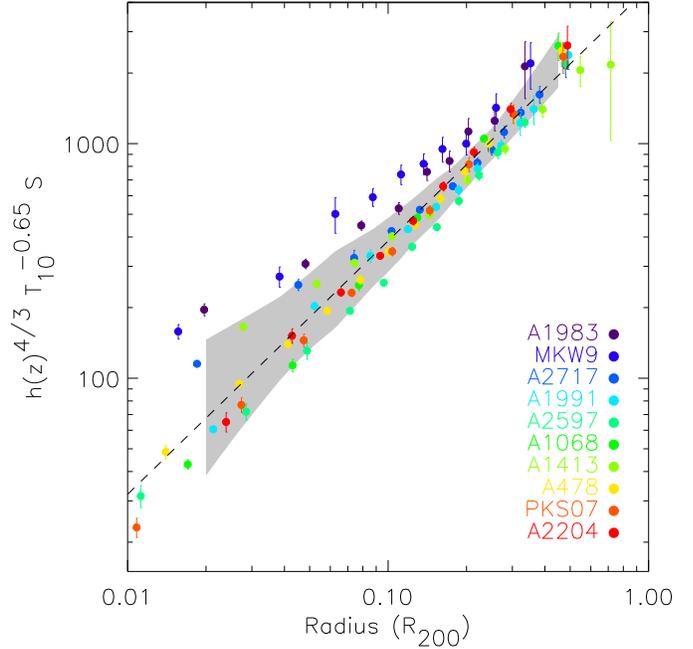}
\caption{{\footnotesize Scaled entropy profiles. The radius is scaled
     to $\rv$ measured from the best-fitting mass models (Papers~I and
     II). The entropy is scaled using the empirical
     entropy scaling $S \propto h(z)^{-4/3} T_{10}^{0.65}$, using
     the global
     temperature, $T$, as listed in Table~\ref{tab:obs}, in units of 10
     keV. The shaded grey area corresponds to the region enclosed by
     the mean plus/minus the $1 \sigma$ standard
     deviation. The dashed line denotes $S \propto R^{1.08}$.}}\label 
{fig:scst}
\end{centering}
\end{figure}

In Fig.~\ref{fig:scst} we show the profiles scaled using the relation
$S \propto h(z)^{-4/3}\,T_{10}^{0.65}$, where $T_{10}$ is the global
temperature 
measured in units of 10 keV. This relation is consistent with our data
(Table~\ref{tab:strel}), and allows us to compare our results with
previous work. As an initial measure of the scatter in scaled entropy
profiles, we estimated the dispersion at various radii in the range
$0.02$--$0.45\,\rv$. The shaded area in Fig~\ref{fig:scst} shows the
region enclosed by the mean plus/minus the $1 \sigma$ standard
deviation.  Figure~\ref{fig:scst} shows that, outside the core
regions, the entropy profiles present a high degree of
self-similarity.  The relative dispersion in scaled profiles
remains approximately constant at $\lesssim 20$ per cent for
$r\gtrsim0.1\,\rv$, in excellent agreement with the dispersion found
in a smaller subsample by \citet{pa05}. In the core regions, however,
the dispersion increases with decreasing radius to reach $\gtrsim
60$ per cent at $\sim 0.02~\rv$.

\begin{figure}[t]
\begin{centering}
\includegraphics[scale=1.,angle=0,keepaspectratio,width=\columnwidth] 
{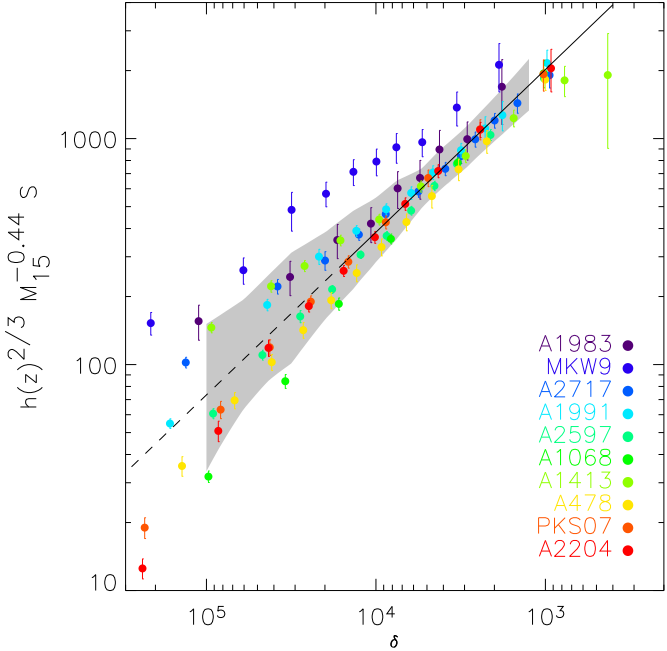}
\caption{{\footnotesize Scaled entropy profiles. The radius is
     expressed as a function of density contrast, $\delta$. The entropy
     has been scaled with the best-fitting entropy-mass scaling $S_ 
{2500}
     \propto h(z)^{-2/3} M_{15}^{0.44}$, using the measured total mass,
     $M_{200}$, in units of $10^{15}$ M$_{\odot}$. The shaded grey area
corresponds to the region enclosed by the mean plus/minus the $1
\sigma$ standard deviation (including MKW9). The solid line shows the
power law relation, $S \propto \delta^{-0.66}$, obtained from fitting
points with $\delta < 15000$ excluding MKW9. The dashed line is an
extrapolation of this best fit. }}\label{fig:smdelta}
\end{centering}
\end{figure}

We next fitted the scaled profiles with a power law using the BCES
method.  The fit was performed in the log-log plane taking into
account the errors on both variables.  Fitting the data in the radial
range $r \ge 0.01\,\rv$ we find a slope of $1.08 \pm 0.04$, with a
dispersion of $\sim 30$ per cent about the best fitting line.  The
slope is not significantly changed ($1.14\pm0.06$) if the data are
fitted in the radial range $r \ge 0.1 \,\rv$, but the intrinsic
dispersion is two times smaller ($14$ per cent).

For comparison, \citet{pa05}, using a subsample of the
current data (A1983, MKW9, A2717, A1991 and A1413), found a slope of
$0.94 \pm 0.14$, while \citet{piff}, using the same scaling, find a
slope of $0.95 \pm 0.02$ for their sample of 13 cool core clusters. In
contrast, broken power-law behaviour has been found by \citet{mad} for
a sample of groups ($\kT \lesssim 2$ keV).
Our slightly higher slope, compared to that of
\citet{piff}, can be explained by a slight difference in the temperature
profile shape. We note simply that the samples are not
equivalent and that a detailed investigation of the slope of the
entropy profile will require a carefully controlled, unbiased
sample spanning the mass range from groups to massive
clusters. Irrespective of the exact slope, that we find $S \propsim 
r^{1.1}$ is interesting
considering that this value is expected for shock heating in spherical
collapse \citep[e.g.,][]{tn01}. This is further discussed in
Sec.~\ref{sec:discuss}. 

\subsection{Scaled entropy versus $\delta$}

In Figure~\ref{fig:smdelta}, we show the entropy profiles scaled using
the relation $S_{2500} \propto h(z)^{-2/3}\,M_{15}^{0.44}$, where $M_
{15}$ is the 
total mass $M_{200}$ in units of $10^{15}$ M$_{\odot}$. Radii are
plotted in terms of the density contrast, $\delta$.  The shaded area
in Fig.~\ref{fig:smdelta} shows the region enclosed by the mean
plus/minus the $1~\sigma$ standard deviation.

Again, while we observe similarity in the entropy profiles outside the
core regions ($\delta \lesssim 15000$), there is a significant
increase in the dispersion at higher density contrast. It is also
clear that MKW9 is a distinct outlier. This may indicate that the mass
of this cluster is underestimated, or its temperature is
overestimated.  Excluding MKW9 and fitting the scaled entropy profiles
with a power law using the BCES method, the slope varies from
$-0.76\pm0.03$ when fitting $\delta < 3 \times 10^5$
($\sim0.01\,\rv$), to $-0.66\pm0.03$, when fitting $\delta < 15000$
($\sim0.1\,\rv$). Thus the entropy does not strictly behave as a
power-law in density contrast.  The intrinsic dispersion drops from
$30$ per cent to $10$ per cent depending on the range of density contrast
fitted. Outside the core regions, the intrinsic dispersion is smaller
than for temperature-scaled profiles, in agreement with the lower
dispersion observed for the $S$--$M$ relation compared to that of the
$S$--$T$ relation. 

\section{Discussion}
\label{sec:discuss}

\begin{figure*}[ht]
\begin{centering}
\includegraphics[scale=1.,angle=0,keepaspectratio,width=0.97 
\columnwidth]{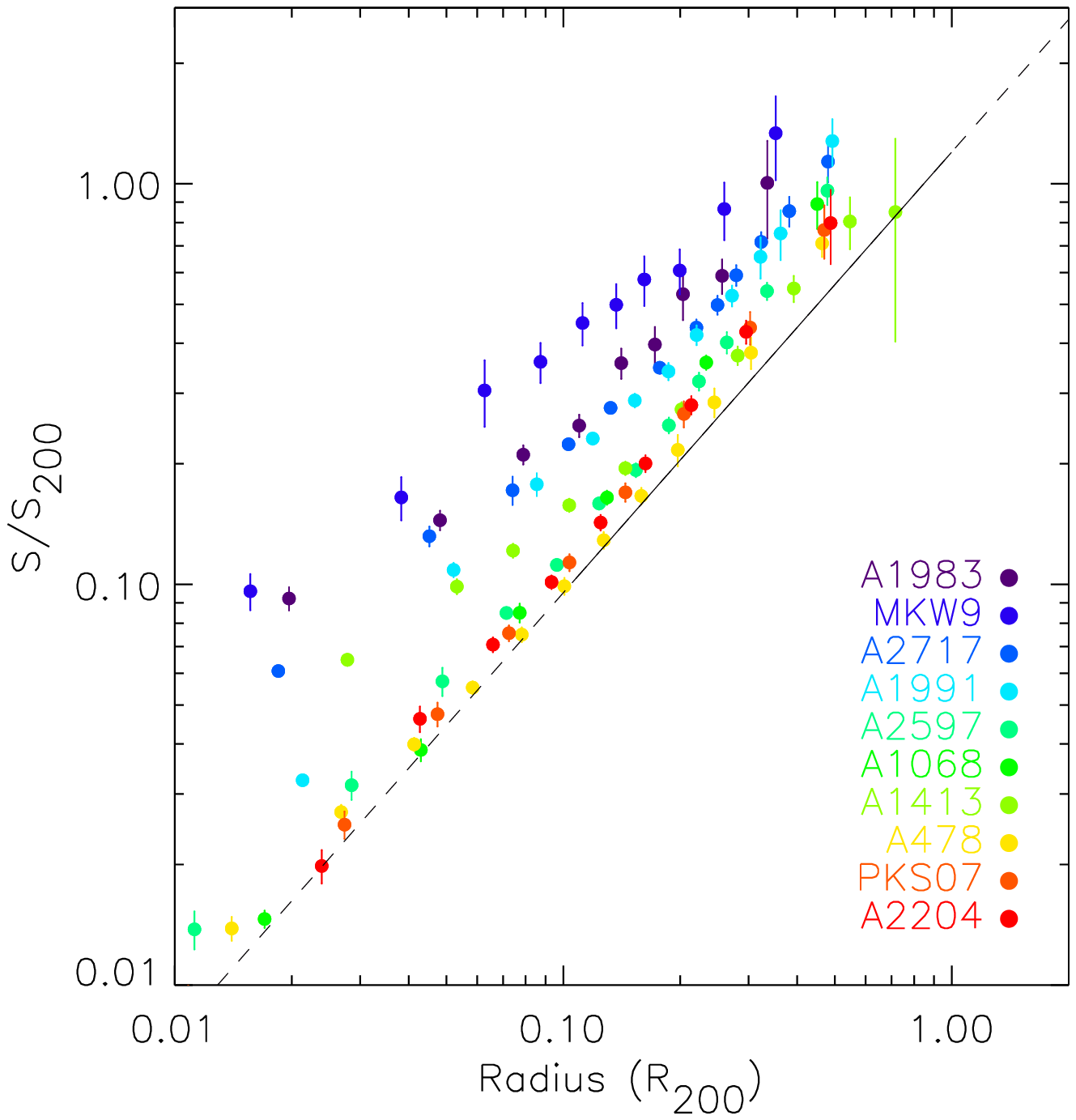}
\hfill
\includegraphics[scale=0.75,angle=0,keepaspectratio,width=0.92 
\columnwidth]{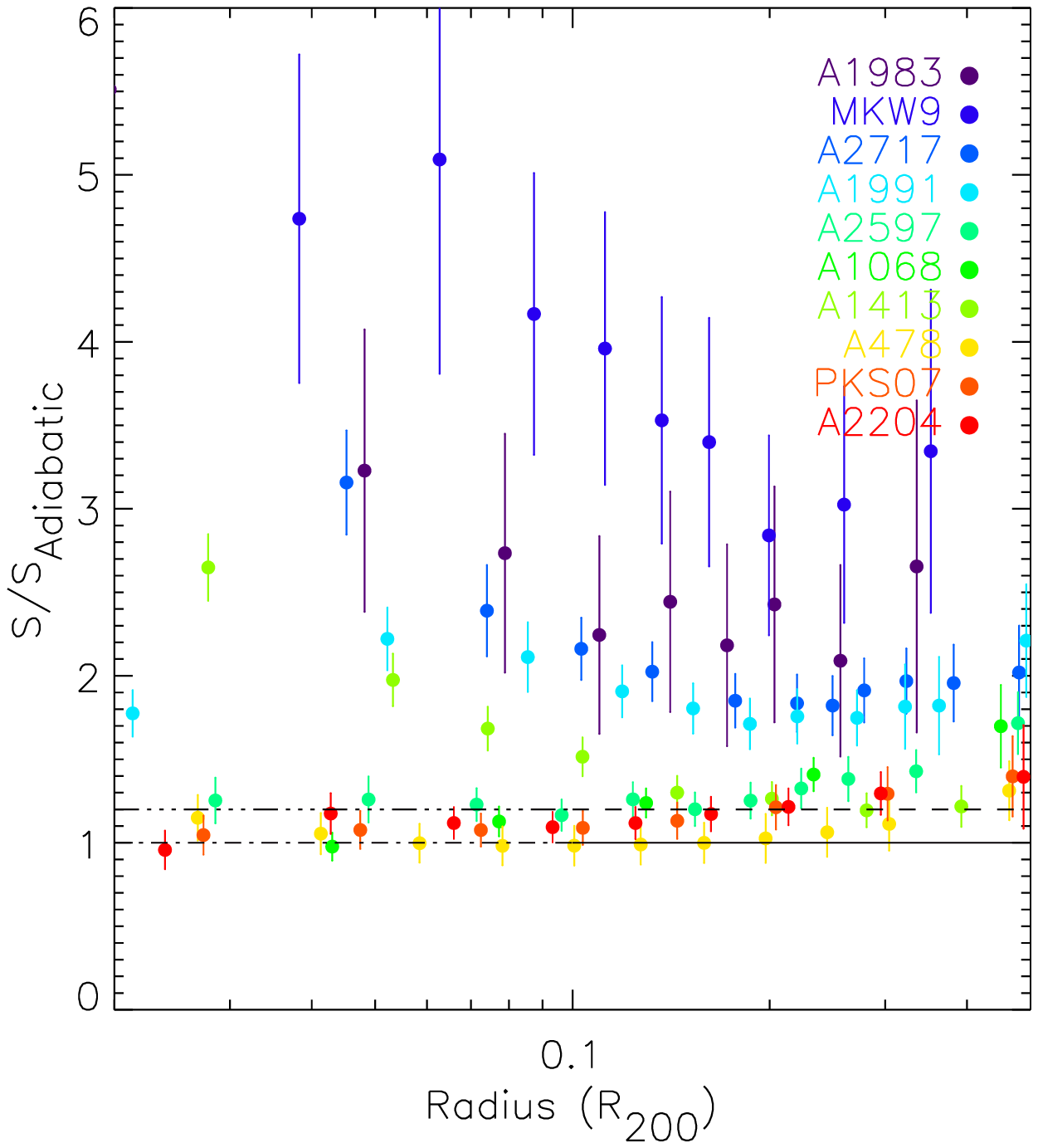}
\caption{{\footnotesize A comparison of our observed entropy profiles
with the prediction from the adiabatic numerical simulations of
\citet{voit05}. {\bf Left panel}: The observed entropy profiles have
been scaled to
$S_{200}$ using Equation~\ref{eqn:voitconv}. The solid line represents
the best-fitting power law relation found by \citet{voit05} from
fitting adiabatic numerical simulations of 30 clusters in the radial
range $0.1 < \rv < 1.0$. {\bf Right panel}: Ratio between
the observed profiles and the best-fitting power law relation found by
\citet{voit05}. The dashed line shows the same 
relation multiplied by 1.2 to take into account the difference of 30
percent between the observed normalistion of the $M$--$T$ relation and
that found in adiabatic simulations.}}\label{fig:compad}
\end{centering}
\end{figure*}

Our results put into evidence two main departures from the standard   
self-similar model of cluster formation. First, beyond the core  
region ($r\gtrsim0.1\,\rv$) the entropy profiles obey self-similarity,
having a shape consistent with expectations, but with a modified  
temperature (or mass) scaling. The scaling relations are shallower  
than expected. Second, there is a break of similarity in the core  
region: the dispersion in scaled profiles increases with decreasing
radius. In this section, we will discuss in turn both characteristics.

\subsection{Entropy normalisation}

The modified entropy scaling indicates that there is an excess of
entropy, in low mass objects {\it relative} to more massive systems,
as compared to the expectation from pure shock heating. A comparison
with adiabatic numerical simulations allows us to quantify the {\it
absolute} value of the excess and to examine whether an entropy excess
is also present for the most massive systems.

\citet{voit05} shows the results of adiabatic numerical
simulations of 30 clusters spanning a mass range of more than a factor
of ten. Once scaled by the characteristic entropy of the halo,

\begin{equation}
S_{200} = \frac{1}{2} \left[\frac{2 \pi}{15} \frac{G^2 M_{200}}{f_b
H(z)}\right]^{2/3},
\end{equation}

\noindent where $f_b$ is the baryon fraction, the simulated profiles
are closely self-similar, and can be well fitted in the radial range
$\sim 0.1 - 1.0~\rv$ by the power-law relation $S/S_{200} =
1.26(R/\rv)^{1.1}$.  Assuming $f_b = 0.14$ ($\Omega_b h^2=0.02$ and
$\Omega_m =0.3$) and typical elemental abundances, we can scale our
observed entropy values to $S_{200}$ using the expression

\begin{equation}
S/S_{200} = \left(\frac{S}{2471~{\rm  keV cm}^2}\right)
\left(\frac{M_{200}}{1\times15 M_\odot}\right)^{-2/3} h(z)^{2/3}. 
\label{eqn:voitconv}
\end{equation}

\noindent Further scaling the radius by the measured value of $\rv$
allows us to compare our results directly with the adiabatic
simulations. In the left hand panel of Figure~\ref{fig:compad}, our
observed entropy profiles are compared with the prediction of
\citet{voit05}. The right hand panel shows the ratio between the
best-fitting adiabatic power-law relation and the observed data.

It can be seen that the richer systems all have entropies which are in
good agreement (both in slope and normalisation) with the adiabatic
prediction, denoted by the solid line in Fig.~\ref{fig:compad}.  On
average, their entropy is only slightly higher than predicted (by
$\sim 20$ per cent), although the effect is not very significant.  We
recall that there is also a $\sim 30$ per cent difference in
normalisation between the observed $M$--$T$ relation and that
predicted by adiabatic simulations \citep{app}. Interestingly, this
corresponds to a $\sim 20$ per cent entropy excess at a given mass for
$T\propto M^{2/3}$.  The (slight) excess of entropy in massive systems
is thus consistent with a simple increase of the mean temperature, i.e.,
of the internal energy of the ICM.  However, Fig.~\ref{fig:compad}
shows explicitly that the poorer systems have a systematically higher
entropy normalisation than the richer systems. There is approximately
2.5 times more entropy at $0.2\,\rv$ in the ICM of A1983, the poorest
cluster in our sample, than that
predicted by gravitational heating.  This excess shows
that the density of the ICM is also affected at lower mass.

\citet{voit03} and \citet{psf03} independently noted that the ICM
entropy is highly sensitive to the density of the incoming gas and
suggested that a smoothing of the gas density due to pre-heating in
filaments and/or infalling groups would boost the entropy production
at the accretion shock.  The entropy amplification is more efficient
in low mass systems, because they accrete smaller halos more affected
by smoothing due to pre-heating.  In this interpretation, no
substantial isentropic core will develop because the amount of initial
preheating needed is substantially less than the characteristic
entropy of the final halo. Clusters will thus be self-similar
down to low mass but with a modified normalisation scaling as
discussed above. This is in agreement with our results outside the
core region ($r \gtrsim 0.1\rv$).

Recent numerical simulations which mimic preheating by imposing a
minimum entropy floor at high $z$ have confirmed the entropy
amplification effect due to smooth accretion
\citep[][]{borg05}. However, the effect seems to be substantially
reduced when cooling is also taken into account. Furthermore, the
physical origin of the preheating is still unclear. Heating by SNe
seems to be too localised to have a significant effect in smoothing
the accreting gas \citep[][] {borg05}.  While AGN might be better
candidates to produce the extra heating, the observed normalisations
require that the AGN affects the entropy distribution at least up to
$r_{1000}$. Some recent theoretical investigations suggest that this
may be possible \citep{roy,vd}. 

\subsection{Localised modification}

While filamentary pre-heating may explain the increased normalisation
of poor systems relative to hotter systems, it does not explain the
increasing scatter towards the central regions in scaled profiles
(Figs.~\ref{fig:scst}, ~\ref{fig:smdelta} and~\ref{fig:compad}). The
adiabatic numerical simulations of \citet{voit05} show both a
flattening of the slope and an increase in the dispersion of the
scaled entropy profiles in the central regions ($< 0.1~\rv$). However,
the dispersion in our observed profiles ($\sim 60$ per cent) greatly
exceeds that of the simulations ($\sim 30$ per
cent, cf Fig.~\ref{fig:compad} and Fig.~11 of \citealt{voit05}).
Six clusters out of our total sample of ten (A1991, A2597, A1068,
A478, PKS0745 and A2204) have remarkably similar scaled entropy
profiles, displaying power-law behaviour down to the smallest radii
measured. Fitting these scaled profiles in the radial range $0.01 -
0.1\,\rv$ with power-law using the BCES method, we find a slope of
$1.13\pm0.05$, with a dispersion of only 13 per cent. This slope is
very similar to that found by fitting the $r > 0.1\,\rv$ range. These
six clusters all appear to host a bona fide cooling core, each having
a central temperature decrement of a factor $\sim 2$
\citep[cf][]{kaa04}. Similar power-law profiles were found by in a
sample of 13 cooling core clusters by \citet{piff}.  Strong radiative
cooling thus appears to generate entropy profiles which display
power-law behaviour down to very small radii (Fig.~\ref{fig:compad}).
This is similar to the quasi-steady-state entropy profile in the models of
\citet[][]{mccarthy}, which include radiative cooling.

The other four clusters in our sample are characterised by a smaller
central temperature decrement, larger cooling times and shallower
entropy profiles. Clearly, some mechanism has modified the cooling
history of these clusters. Energy input from AGN is regularly invoked
as a way of moderating cooling at the centres of galaxy clusters
\citep[e.g.,][and references therein]{churazov}. 
Our sample contains four clusters which have X-ray evidence for
interactions between radio and X-ray plasma (A478, A2204, A2597, and
PKS0745), and yet the entropy profiles of all of these clusters
increase monotonically outward in the canonical fashion. However, we
note that {\it Chandra} observations of clusters with moderate to
strong radio sources (including A478), show some evidence for a
slightly shallower slope ($\sim 0.8-0.9$) but only in the very inner
regions \citep*[$R \lesssim 0.05~\rv$; see
e.g.,][]{belsole,johnstone,sfm}. 
The effect of AGN activity on the entropy profile will depend on
whether the energy input is catastrophic (i.e., occurring in strong
bursts) or distributed (more moderated input). It
is clear that if the heating is catastrophic in nature, no cluster has
yet been found in which this is evident at least from the point of
view of the entropy (although see \citealt{mcnamara,nulsen}). The
heating is likely more distributed, via e.g., weak shocks
\citep[][]{fabian03}, thus preserving the generally increasing form of
the entropy profile. 

Merging events can result in substantial mixing of high and low
entropy gas \citep[e.g.][]{rs01}. Relatively little attention has been
given to this possibility in the literature (see, however, Fig.~5 of
\citealt{voit03}, Fig.~12 of \citealt{belsole04} and Fig.~6 of
\citealt*{pbf}). Such redistribution of entropy will depend on the
scale of the merger, whether the merger has disrupted the structure of
the cool core, and the timescale for re-establishment of the cool core
if disrupted. In the current sample MKW9, A2717, A1413 and A1983 all
have flatter core entropy profiles.
We note that the morphological information for the present
sample would argue against recent merger activity in these
clusters (\citealt{pa05}; Papers~I and~II). However, this does not
rule out entropy modification due to a more ancient merger,
particularly if the relaxation time is less than the cooling time
\citep[see also][]{belsole04}. 


\begin{table*}
\centering

\begin{minipage}{\textwidth}
\center
\caption{{\footnotesize Columns: (1) Cluster name; (2) Parametric
   model; (3) Central density
   of principal $\beta$
model ($\times 10^{-2}$ $h_{70}^{1/2}$ cm$^{-3}$); (4,5) Central  
density of
second and third $\beta$ model components ($\times 10^{-2}$ $h_{70}^ 
{1/2}$
cm$^{-3}$) (6) Core radius of inner $\beta$-model (arcmin); (7) Core
radius of principal $\beta$-model (arcmin); (8,9) Core radius of
second and third $\beta$ model components (arcmin); (10) $\beta$
parameter; (11) Radius of boundary between $\beta$ model components;
(12) $\xi$ parameter; (13) $\chi^2$/d.o.f. }}
\begin{tabular}{l l l l l l l l l l l l l l l}

\hline
\hline
\multicolumn{1}{l}{ Cluster } & \multicolumn{1}{l}{Model}
& \multicolumn{1}{l}{$n_{e_1}$}
& \multicolumn{1}{l}{$n_{e_2}$} & \multicolumn{1}{l}{$n_{e_3}$}
& \multicolumn{1}{l}{ $r_{c,{\rm in}}$ } & \multicolumn{1}{l}{
   $r_{c_1}$ }
& \multicolumn{1}{l}{$r_{c_2}$} & \multicolumn{1}{l}{$r_{c_3}$}
& \multicolumn{1}{l}{ $\beta$ } & \multicolumn{1}{l}{$r_{\rm cut}$}
& \multicolumn{1}{l}{$\xi$} & \multicolumn{1}{l}{$\chi^2$/d.o.f.} \\

\hline

A1983   & BB    & 0.64  & --    & --    & 0.49  &
2.65  & --    & --    & 0.76 & 2.40  & 1.    & 71/64  \\

MKW9    & BB    & 0.79  & --    & --    & 0.35  &
3.36  & --    & --    & 0.70 & 3.40  & 1.    & 181/133  \\

A2717   & BB    & 1.23  & --    & -     & 0.31  &
1.96  & --    & --    & 0.63 & 2.08  & 1.    & 339/206  \\

A1991   & BB    & 5.61  & --    & -     & 0.16 &
1.42  & --    & --    & 0.65 & 2.08 & 1.    & 197/160  \\

A2597   & KBB   & 4.30  & --    & -     & 0.32  & 1.21
& --    & --    & 0.73 & 1.92  & 2.30  & 158/125  \\

A1068   & KBB   & 40.31 & --    & -     & 0.58  & 3.20
& --    & --    & 1.01 & 4.53  & 0.25  & 58/42  \\

A1413   & KBB   & 3.66  & --    & --    & 0.41  &
1.34  & --    & --    & 0.71 & 1.69 & 0.50 & 65/47  \\

A478    & 3B    & 9.74  & 2.33  & 0.36  & --    &
0.25 & 1.12 & 3.30 & 0.84 & --    & --    & 50/56 \\

PKS0745 & KBB   & 18.26 & --    & --    & 0.19  & 0.81
& --    & --    & 0.64 & 1.34  & 0.55  & 260/214  \\

A2204   & 3B    & 15.18 & 1.42  & 0.13  & --    & 0.20
& 1.09  & 3.87  & 0.91 & --    & --    & 38/44 \\

\hline
\end{tabular}
\label{tab:sbres}
\end{minipage}
\end{table*}

\section{Conclusions}

The present sample of ten relaxed clusters constitutes the first with
precise mass data and wide temperature coverage, allowing the detailed
study of the scaling of ICM entropy with both temperature and
mass. The entropy profiles are sampled with good spatial
resolution up to $\sim 0.5\,\rv$, allowing us to examine structural
properties avoiding the stacking
analysis and extrapolation on which previous {\it ROSAT/ASCA} works
relied. 

We have found that the entropy profiles of the present cluster sample
are self-similar beyond $\sim 0.1\,\rv$, with a shape not
significantly different from that expected from shock heating.  The
entropy scaling relations are shallower than expected from standard
self-similar models and adiabatic numerical simulations. The slopes of
the entropy scaling relations are independent of radius, reflecting
the structural self-similarity. The entropy-mass (\sm) relation is
consistent with the observed entropy-temperature (\st) and
mass-temperature ($M$--$T$) relations. The dispersion is smaller about
the \sm\ relation, reflecting the fact that the mass of a cluster is
its most fundamental characteristic. These results confirm the trends
seen in smaller samples \citep{pa05}, and in stacking analysis of $\it
ROSAT/ASCA$ data \citep{psf03}. 

Comparison of our observed profiles with adiabatic numerical
simulations has allowed us to quantify the entropy excess relative to
that expected in pure gravitational collapse. The excess is only $\sim
20$ per cent in our highest-mass systems. This is in line with a
simple increase in the mean temperature consistent with the observed
difference in normalisation between the observed $M$--$T$ relation and
that predicted by adiabatic simulations. However, the excess in low
mass systems can be of more than a factor of 2.5, indicating that the
density of the ICM is also affected in these systems. These characteristics are
in qualitative agreement with a scenario in which entropy production
is boosted at the accretion  
shock. A plausible candidate mechanism is smoothing of the accreted 
gas due to preheating \citep{voit03,psf03}.

We have found new and compelling evidence for an increase in the
entropy dispersion in the core regions $\sim 60$ per cent at
$0.02\,\rv$. However the five cooling core 
clusters in our sample have remarkably self-similar power-law profiles with a
dispersion of only 13 per cent between 0.01 and $0.1\,\rv$. The
observed increase in dispersion towards the central regions argues 
for localised entropy modification mechanisms. We conclude that 
AGN activity and/or transient entropy modification due to merging
events are good candidates for the modification of cooling in these
clusters. 

These will be interesting questions to address with
numerical simulations. On the observational side, a larger, unbiased,
sample of clusters will undoubtedly provide greater insights. 

\begin{acknowledgements}

GWP thanks E. Belsole and J.P. Henry for useful discussions, and
acknowledges funding from a Marie Curie Intra-European Fellowship
under the FP6 programme (Contract No. MEIF-CT-2003-500915). EP
acknowledges the financial support of CNES, the French Space Agency,
and of the Leverhulme trust (UK). The present work is based on
observations obtained with {\it XMM-Newton}, an ESA science mission
with instruments and contributions directly funded by ESA Member
States and the USA (NASA).

\end{acknowledgements}

\appendix

\section{Surface brightness profile fit results}

The co-added EPIC azimuthally averaged, vignetting corrected,
background subtracted surface brightness profile was computed in the
[0.3-3.] keV band for each cluster. The profile was corrected for
radial variations in the emissivity due to abundance or temperature
gradients as described in \citet{pa03}. The gas density profile was
obtained from fitting parametric analytic models, convolved with the
EPIC PSF, to the corrected profile. As discussed in Paper~I, we tried
various models and empirically chose the best fitting model using the
$\chi^2$ statistic as a measure of the goodness of fit. The models
used were: (i) the double $\beta$ (BB) density model \citep[] 
[Eq.~3--5]{pa02}; (ii) the
modified double $\beta$ (KBB) model, allowing a more concentrated gas
density distribution towards the centre \citep[][Eq.~6--8]{pa02}, and
(iii) the sum of
three $\beta$ (3B) models for the {\it emission measure}, in which a  
common value of $\beta$ is used
to ensure smooth behaviour at large radii
\citep[][Eq.~2--3]{pakp}.

\noindent The BB and KBB models can be written:

\begin{equation}
\begin{array}{lllll}

r < \rcut & n_e(r)& = &n_{e_1} &\left[ 1 +
\left(\frac{r}{\rci}\right)^{2\xi}\right]^{-\frac{3 \beta_{\rm
in}}{2\xi}} \\

r > \rcut & n_e(r)& = & N &\left[ 1 + \left(\frac{r}{\rc}\right)^2
\right]^{-\frac{3 \beta}{2}},\\
   \end{array}
\end{equation}

\noindent where $\xi = 1$ for the BB model and $\xi < 1$ for the KBB
model, and 

\begin{equation}
N = n_{e_1} \frac{ \left[ 1+ \left( \frac{\rcut}{\rci} \right)^{2\xi}
\right]^{\frac{-3\beta_{\rm in}}{2\xi} } } { \left[ 1+ \left(
\frac{\rcut}{\rc}\right)^{2} \right]^{-\frac{3\beta}{2} } }
\end{equation}

\noindent with

\begin{equation}
\beta_{\rm in} = \beta~\frac{1
+\left(\frac{\rci}{\rcut}\right)^{2\xi}}
{1  +\left(\frac{\rc}{\rcut}\right)^{2}}.
\label{eq:betai}
\end{equation}
  
\noindent The density profile of the BBB model can be written:

\begin{equation}
n_e(r) = \sqrt{n_{e_1}^2 (r) + n_{e_2}^2 (r) + n_{e_3}^2 (r)}, 
\end{equation}

\noindent with

\begin{equation}
n_{e_i}(r) = n_{e_i} \left[ 1 + \left(\frac{r}{r_{c_i}}\right)^2 \right]^{-\frac{3\beta}{2}} .
\end{equation}

\noindent The best-fitting models, and model parameters, are listed in
Table~\ref{tab:sbres}.

\end{document}